\documentclass{aa}  

\usepackage{graphicx}
\usepackage{txfonts}
\usepackage{lipsum}
\usepackage{lscape}             % to rotate a single page table, example in appendix.
\usepackage{placeins}           % useful with \FloatBarrier, to keep 
\usepackage{natbib}
\usepackage{xcolor}
\usepackage[dvipsnames]{xcolor}
\usepackage{multirow}
\usepackage{multicol}
\usepackage{makecell}
\usepackage{breakcites}
\usepackage[pdfencoding=auto,psdextra]{hyperref}
\hypersetup{
    colorlinks=true,
    linkcolor=blue,
    filecolor=magenta,      
    urlcolor=blue,
    citecolor=blue
}
\begin{document}

%%%%%%%%%%%%%%%%%%%%%%%%%%%%%%%%%%%%%%%%
% if you use custom commands in your title,
% ensure to check your title when submitting!
%%%%%%%%%%%%%%%%%%%%%%%%%%%%%%%%%%%%%%%%
   \title{Impact of source clustering on weak lensing higher-order statistics}

%%%%%%%%%%%%%%%%%%%%%%%%%%%%%%%%%%%%%%%%
% Please separate each author with the \and command
%
% Please do not include ORCIDs next to author names.
% Only ORCIDs authenticated by individual authors in EDPS
% editorial system will be taken into account.
% ORCIDs included here will be removed.
%%%%%%%%%%%%%%%%%%%%%%%%%%%%%%%%%%%%%%%%

   \author{Seung-gyu Hwang\inst{1}\fnmsep\thanks{seunggyu.hwang@cea.fr}
        \and Natalia Porqueres\inst{1}
        \and Andreas Tersenov \inst{3,4}
        \and Jean-Luc Starck\inst{1,2}
        }

   \institute{Université Paris-Saclay, Université Paris Cité, CEA, CNRS, AIM, 91191, Gif-sur-Yvette, France
   \and Institutes of Computer Science and Astrophysics, Foundation for Research and Technology Hellas (FORTH), Greece
   \and Department of Astronomy, University of Virginia, 530 McCormick Road, Charlottesville, VA 22904, USA
   \and Virginia Institute for Theoretical Astronomy, University of Virginia, Charlottesville, VA 22904, USA}

   \date{Received November 27, 2026}

% \abstract{}{}{}{}{}
% 5 {} token are mandatory
 
\abstract
    {
    Weak gravitational lensing provides a powerful probe of the large-scale structure of the Universe and the distribution of dark matter. Upcoming Stage IV surveys will deliver precise measurements of billions of galaxies, motivating the development of higher-order statistics to extract the non-Gaussian information encoded in the weak lensing signal. However, these statistics can be sensitive to observational and astrophysical systematics. In particular, source-galaxy clustering induces correlations between the lensing signal and shape noise, potentially biasing cosmological inference. \\
    In this work, we investigate the impact of source clustering on higher-order weak lensing statistics and quantify its consequences for cosmological parameter estimation. Using the Gower Street simulation suite, we generate convergence maps with and without source clustering and analyse them using three summary statistics: third-order moments, wavelet peak counts, and the wavelet $\ell_1$-norm. We employ simulation-based inference to derive cosmological parameter posteriors and assess the impact of source clustering on parameter biases and the figure of merit. \\
    We find that source clustering has its largest impact at low redshifts and on small angular scales. When source clustering is present in the data but neglected in the inference model, cosmological parameter estimates can be biased by up to $3.3\sigma$. Among the statistics considered, third-order moments exhibit the greatest robustness to source clustering, whereas the wavelet $\ell_1$-norm provides the highest figure of merit. These findings reveal an important trade-off between constraining power and robustness to model misspecification: summary statistics that provide tighter cosmological constraints may also be more vulnerable to unaccounted-for systematics. Accounting for source clustering will therefore be important to fully exploit the statistical potential of higher-order weak lensing analyses in Stage IV surveys.
    }
    
    \maketitle

%%%%%%%%%%%%%%%%%%%%%%%%%%%%%%%%%%%%%%%%%%%%%%%%%%%%%%%%%%%%%%
\section{Introduction}
\label{sect:introduction}

Weak gravitational lensing arises from the deflection of light by the gravitational potential of foreground large-scale structures, leading to coherent distortions in the observed shapes of background galaxies. By measuring these distortions across large galaxy samples, cosmic shear provides a powerful cosmological probe, and it is one of the main observables for upcoming Stage-IV cosmological surveys, including \textit{Euclid} \citep[Euclid Consortium;][]{EuclidSkyOverview}, the Vera C. Rubin Observatory \citep{Ivezic2019}, and the Roman Space Telescope \citep{Spergel2015}.

Weak lensing analyses typically assume that source galaxies are spatially uniformly distributed within the survey footprint. However, source galaxies trace the underlying matter density field, and their spatial distribution correlates with the local matter structures. This effect, known as source clustering or source-lens clustering \citep{Bernardeau98, Hamana2002}, modifies the observed weak lensing signal and might bias cosmological parameter inference.

The impact of source clustering on cosmic shear two-point statistics has been investigated in several studies. Using $N$-body simulations, \cite{Yu25} quantified its effect on the lensing power spectra and found deviations of $1$--$10\%$, considering only two tomographic bins with simplifying assumptions. \cite{krause2021darkenergysurveyyear} and \cite{Linke25} extended this to Stage-III survey configurations with realistic redshift distributions corresponding to the Dark Energy Survey \citep[DES,][]{Abbott2016} and the Kilo-Degree-Survey-1000 \citep[KiDS,][]{Asgari2021}, respectively. They quantified the bias induced by source clustering in the $S_8$ and $\Omega_m$ two-point statistics for cosmic shear. Both studies found that when nuisance parameters are accounted for and sampled, the impact of source clustering on cosmological constraints is negligible: the shifts induced by source clustering are absorbed by changes in the intrinsic alignment parameters and in the source redshift distribution. When the nuisance parameters are fixed, the source clustering produces shifts of $0.42\sigma$ in $S_8$ and $0.31\sigma$ in $\Omega_m$.

For a \textit{Euclid}-like setup, source clustering was predicted to induce biases of the order of $1\sigma$  \citep{Deshpande-EP28}. However, these estimates were obtained using a Fisher-matrix formalism, which relies on a local Gaussian approximation to the likelihood and underestimates the posterior width, leading to tighter constraints. 

Source clustering can have a larger impact on higher-order weak lensing observables, which probe non-Gaussian features of the matter distribution. Recent studies \citep{Ajani-EP29, EP-Vinciguerra, Makinen2025, Williamson2026, tersenov2026mitigatingbaryoniceffectsweak} have found that higher-order statistics can provide up to a factor of 3 improvement in the precision of cosmological parameters compared with two-point estimators. However, their increased precision also implies greater sensitivity to modelling uncertainties. 

For three-point statistics, \cite{Valageas14} found that source clustering can account for up to $10\%$ of the weak lensing signal. Similarly, \cite{Bernardeau98} showed that the skewness and kurtosis of convergence maps have a similar sensitivity to source clustering, with a correction term of up to $30\%$ for broad source redshift distributions. The impact of this effect depends strongly on the redshift distribution of the sources \citep{Hamana2002} and decreases significantly for narrow redshift distributions \citep{Bernardeau98}. More recently, \cite{Gatti24} measured the source clustering effect on higher-order weak lensing statistics in DES-Y3, finding that third moments and wavelet phase harmonics are more strongly affected than peak counts. They also introduced a mitigation model based on linear galaxy bias, which was then applied in \cite{Gatti2024_sbi}. Nevertheless, the extent to which source clustering biases cosmological constraints from higher-order weak lensing statistics has not yet been quantified.

In this work, we quantify the impact of source clustering on cosmological parameter constraints from non-Gaussian weak lensing statistics. We consider third-order moments, the wavelet $\ell_1$-norm, and wavelet peak counts in a \textit{Euclid}-like configuration. Using simulation-based inference, we quantify biases in cosmological constraints and assess the sensitivity of these observables to source clustering.

%%%%%%%%%%%%%%%%%%%%%%%%%%%%%%%%%%%%%%%%%%%%%%%%%%%%%%%%%%%%%%
\section{Data and simulations}
\label{sect:data_and_simulations}
\subsection{Gower Street simulation suite}
\label{sect:gower_street}
We use the Gower Street simulation suite \citep{Jeffrey2025_GowerSt} to investigate the impact of source clustering on weak lensing higher-order statistics and the resulting cosmological constraints.
The suite consists of 791 gravity-only full-sky N-body simulations generated with the \texttt{PKDGRAV3} code \citep{Potter2017_PKDGRAV3}.
The simulations span a seven-dimensional parameter space in $w$CDM cosmology, defined by $\Omega_\mathrm{m}, \sigma_8, n_s, h, \Omega_\mathrm{b}h^2, w, \text{ and }m_\nu$.

Each simulation is constructed from replicated boxes with a side length of $L=1250h^{-1}\mathrm{Mpc}$, each containing $N=1080^3$ particles.
Up to ten replications in each direction are used to construct light cones spanning the redshift range $(0\leq z \leq 49)$.
For each simulation, approximately 100 lens planes are provided at different redshifts, equally spaced in proper time.
These lens planes are supplied as HEALPix maps with $N_\mathrm{SIDE}=2048$, encoding the shell overdensity field $(\delta_{\rm shell}(\hat{\boldsymbol n}, \chi))$.

\subsection{Simulated convergence maps}
\label{sect:pipeline_maps}
In this work, we use 599 of the 791 simulations in the Gower Street suite, excluding the initial 192 verification runs. We retain the original HEALPix resolution of \(N_{\mathrm{SIDE}}=2048\) without downsampling to preserve information on small angular scales, where the impact of source clustering is expected to be stronger \citep{Gatti24}.

\begin{figure}[ht!]
    \centering
    \includegraphics[width=80mm]{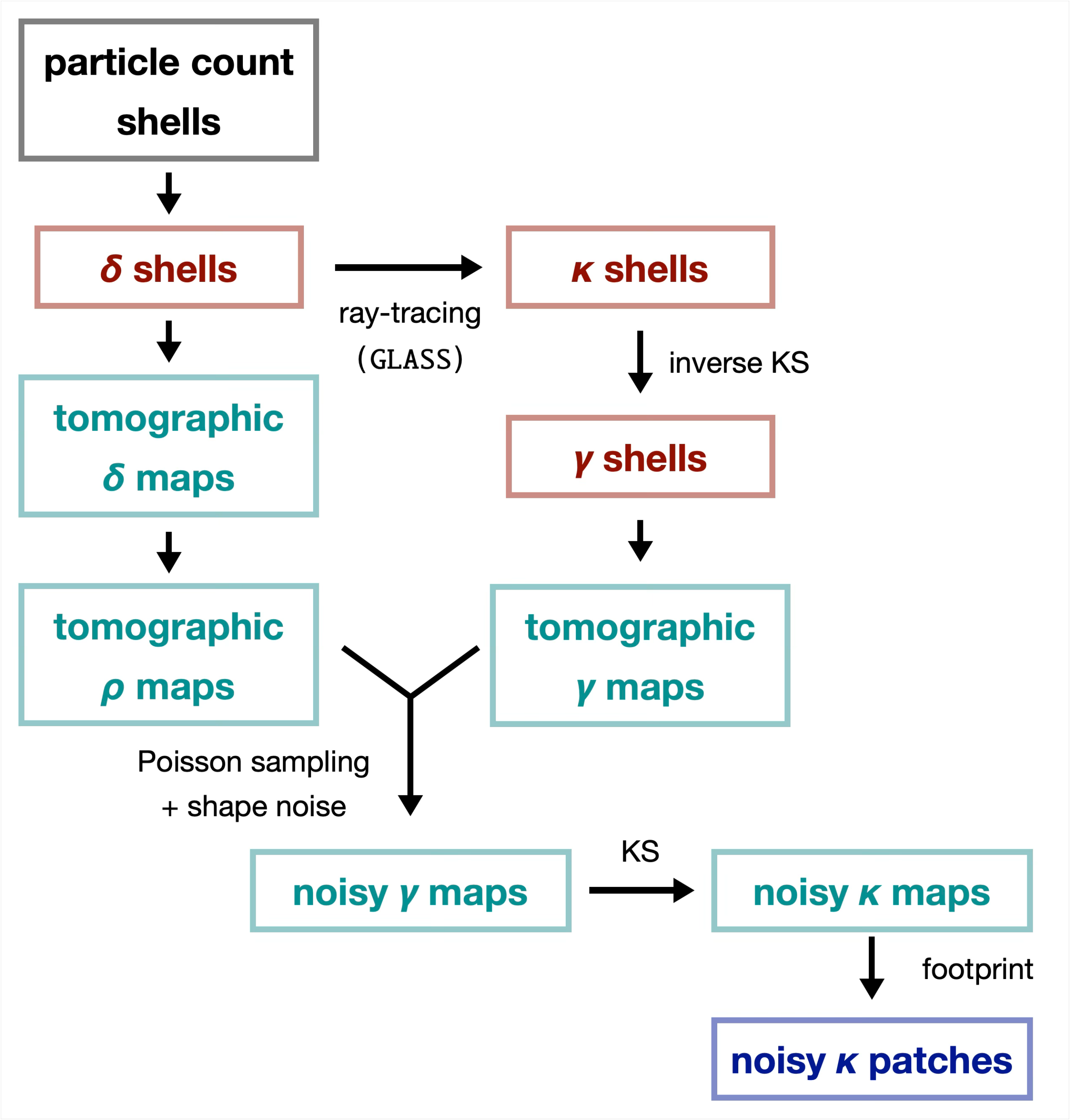}
    \caption{Pipeline for constructing the simulated convergence maps. Colours distinguish shell maps (dark red), tomographic maps (teal), and final convergence patches (dark blue). We use $\delta$, $\rho$, $\kappa$, and $\gamma$ for the matter overdensity, relative source density, weak lensing convergence, and shear, respectively.}
    \label{fig01:pipeline}
\end{figure}

Fig.~\ref{fig01:pipeline} summarises the pipeline used to construct the mock convergence maps employed in this work.
Starting from the raw particle count shells of the Gower Street simulations, we derive the corresponding matter overdensity shells, $\delta_{\mathrm{shell}}(\hat{\boldsymbol{n}},\chi)$, where $\hat{\boldsymbol{n}}$ indicates the line-of-sight direction and $\chi$ is the comoving distance.
These overdensity shells are then used to generate both the lensing signal and the source galaxy density field.
When constructing the tomographic shear maps, the source distribution is assumed to be uniform in the absence of source clustering.
With source clustering, it follows the underlying matter density, changing the relative contribution of each redshift shell at each pixel and thereby affecting the local galaxy counts and, hence, the shape noise.

For the lensing branch of the pipeline, we use the ray tracing implementation in \texttt{GLASS} \citep{Tessore2023_GLASS} to convert the matter overdensity shells into convergence shells, $\kappa_{\rm shell}(\hat{\boldsymbol{n}},\chi)$.
The corresponding shear shells, $\gamma_{\rm shell}(\hat{\boldsymbol{n}},\chi)$, are then derived using the inverse Kaiser--Squires (KS) transformation \citep{Kaiser1993_KS}.
We set the maximum multipole to $\ell_{\max}=3N_{\mathrm{SIDE}}-1=6143$, following the standard HEALPix convention.
The shear shells are subsequently combined according to the source redshift distribution to construct the tomographic shear maps.

We adopt the Stage IV-like observational configuration described in \cite{Castiblanco2024_kpdf}, including the source redshift distribution, galaxy number density, and shape noise properties.
The source redshift distribution extends to $z_{\rm s}\sim 3$ and is divided into five equipopulated tomographic bins, which span the redshift ranges $[0.00,0.47)$, $[0.47,0.72)$, $[0.72,0.97)$, $[0.97,1.34)$, and $[1.34,3.00]$, respectively.

We define the relative source density maps using a linear galaxy bias model as
$\rho(p,s)\equiv 1+b_{\rm g}\delta(p,s)$,
where $\delta(p,s)$ is the matter overdensity at pixel $p$ and redshift shell $s$.
Although more realistic galaxy bias models are available, we adopt this simple prescription to isolate the impact of source clustering.
At first order, both the lensing signal weighting and the shape noise normalisation have a linear dependence on the galaxy bias $b_{\rm g}$.
We therefore expect the impact of source clustering to increase with $b_{\rm g}$, at least for sufficiently small values.
\cite{Gatti24} reported a stronger impact of source clustering for larger galaxy bias values, particularly for $b_{\rm g}>1$.
However, the response of the higher-order summary statistics may not follow a strictly linear scaling due to their nonlinear dependence on the convergence field.

Throughout this work, we consider two configurations: $b_{\rm g}=0$ and $b_{\rm g}=1$, corresponding to the cases without and with source clustering, respectively.
For these two configurations, the resulting tomographic shear map is given by
% \begin{equation}
%     \label{eq00_shear_sc}
%     \begin{aligned}
%     \gamma(p) ={}&
%     \frac{\sum_s \bar{n}(s)\rho(p,s)\gamma(p,s)}
%     {\sum_s \bar{n}(s)\rho(p,s)}
%     \\
%     &+
%     \left(
%     \frac{\sum_s \bar{n}(s)}
%     {\sum_s \bar{n}(s)\rho(p,s)}
%     \right)^{1/2}
%     F_\mathrm{SC}(p)\frac{\sum_g w_g e_g}{\sum_g w_g},
%     \end{aligned}
% \end{equation}
\begin{equation}
    \begin{aligned}
    \gamma_i(p) ={}&
    \frac{\sum_{s\in\mathcal{S}_i} \bar{n}(s)\rho(p,s)\gamma(p,s)}
    {\sum_{s\in\mathcal{S}_i} \bar{n}(s)\rho(p,s)}
    \\
    &+
    \left(
    \frac{\sum_{s\in\mathcal{S}_i} \bar{n}(s)}
    {\sum_{s\in\mathcal{S}_i} \bar{n}(s)\rho(p,s)}
    \right)^{1/2}
    F_\mathrm{SC}(p)\frac{\sum_g w_g e_g}{\sum_g w_g},
    \end{aligned}
    \label{eq01:tomographic_shear}
\end{equation}
where $\mathcal{S}_i$ denotes the redshift shells included in tomographic bin $i$, $\bar{n}(s)$ is the mean source density in shell $s$, $\gamma(p,s)$ is the shear at pixel $p$ from shell $s$, and the sums over $g$ run over galaxies in pixel $p$, with $w_g=1$ and intrinsic ellipticities $e_g\sim\mathcal{N}(0,\sigma_\epsilon^2)$.
The first term describes the lensing signal, weighted by the local source density, while the second term accounts for shape noise, with a normalisation that depends on the source density.
The factor $F_\mathrm{SC}(p)$ corrects for the enhancement of the even moments of the shape noise induced by source clustering and mildly depends on the underlying cosmology.
Since our analysis spans multiple cosmologies, we do not apply this correction and set $F_\mathrm{SC}(p)=1$ throughout.
This approximation neglects a small correction that reduces the shape noise by at most approximately 3\%.

We use the relative source density maps to generate mock galaxy counts via Poisson sampling and to estimate the noise standard deviation at each pixel.
We assign Gaussian shape noise to individual galaxies, using an intrinsic ellipticity dispersion of $\sigma_\epsilon=0.26$ and a galaxy number density of $n_g=\sum_s\bar{n}(s)=6\,\mathrm{arcmin}^{-2}$ per tomographic bin, corresponding to a total of $30\,\mathrm{arcmin}^{-2}$ across all five bins.

We apply the KS transformation to the noisy tomographic shear maps to reconstruct the convergence field.
We then extract 14 non-overlapping patches for each simulation centred at different sky coordinates and project them onto flat-sky maps, each covering $10\,\mathrm{deg}\times10\,\mathrm{deg}$.
We treat the 14 patches as approximately independent samples on the angular scales probed by our summary statistics, while capturing spatial variations at fixed cosmology.
However, residual correlations between patches may persist because large-scale modes extend beyond individual patch boundaries, potentially reducing the effective number of independent samples and leading to an underestimation of the scatter in our results.
This results in a total of 8,386 convergence patches across the 599 cosmological simulations, which are used in the subsequent analysis.

\begin{figure*}[ht!]
\centering
\includegraphics[width=170mm]{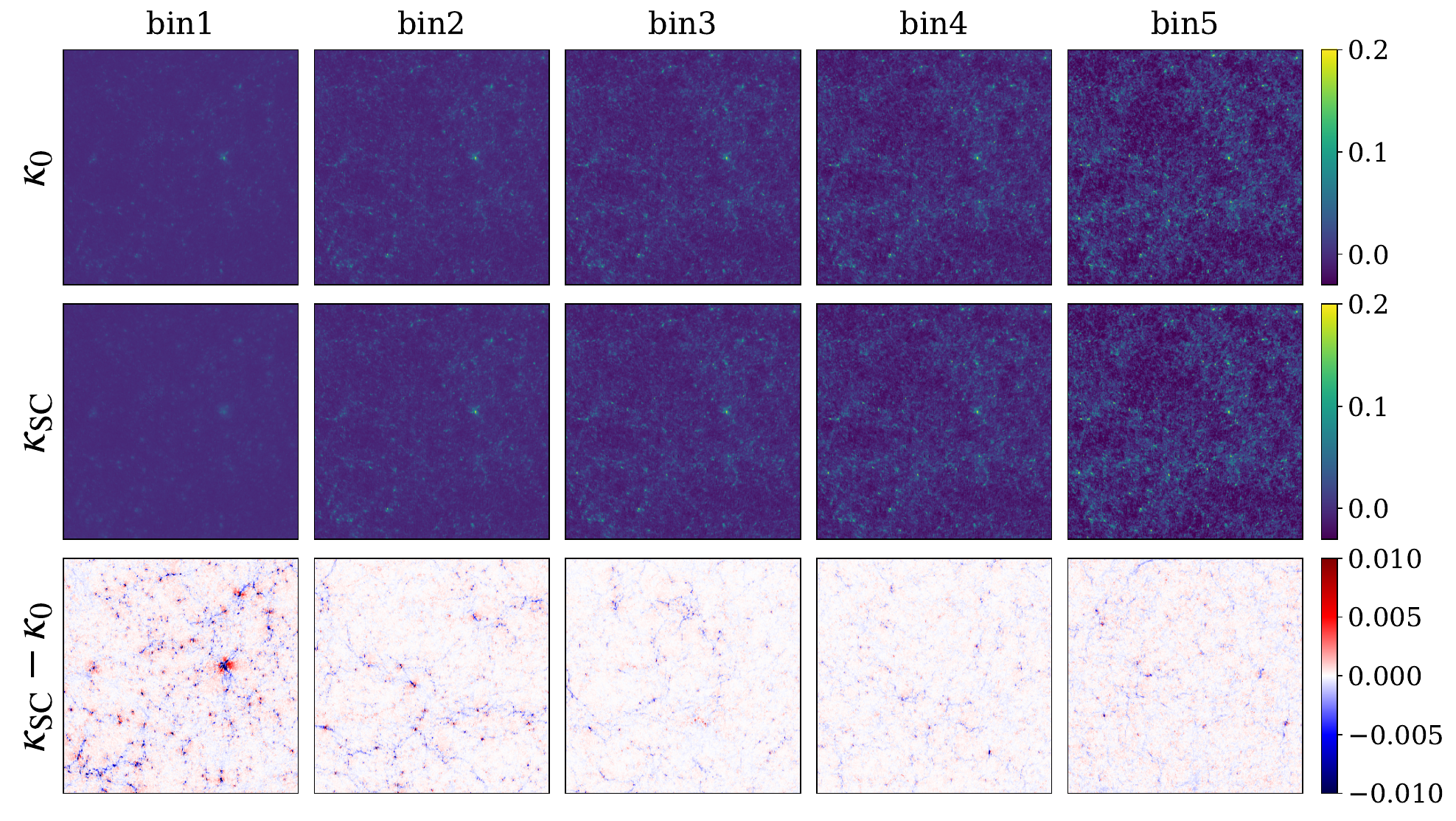}
    \caption{Noiseless convergence maps for the five tomographic bins. The first and second rows show the maps without source clustering ($\kappa_0$) and with source clustering ($\kappa_{\mathrm{SC}}$), respectively. The third row shows their difference ($\kappa_{\mathrm{SC}}-\kappa_0$).}
\label{fig02:convergence_noiseless}
\end{figure*}

\begin{figure*}[ht!]
\centering
\includegraphics[width=170mm]{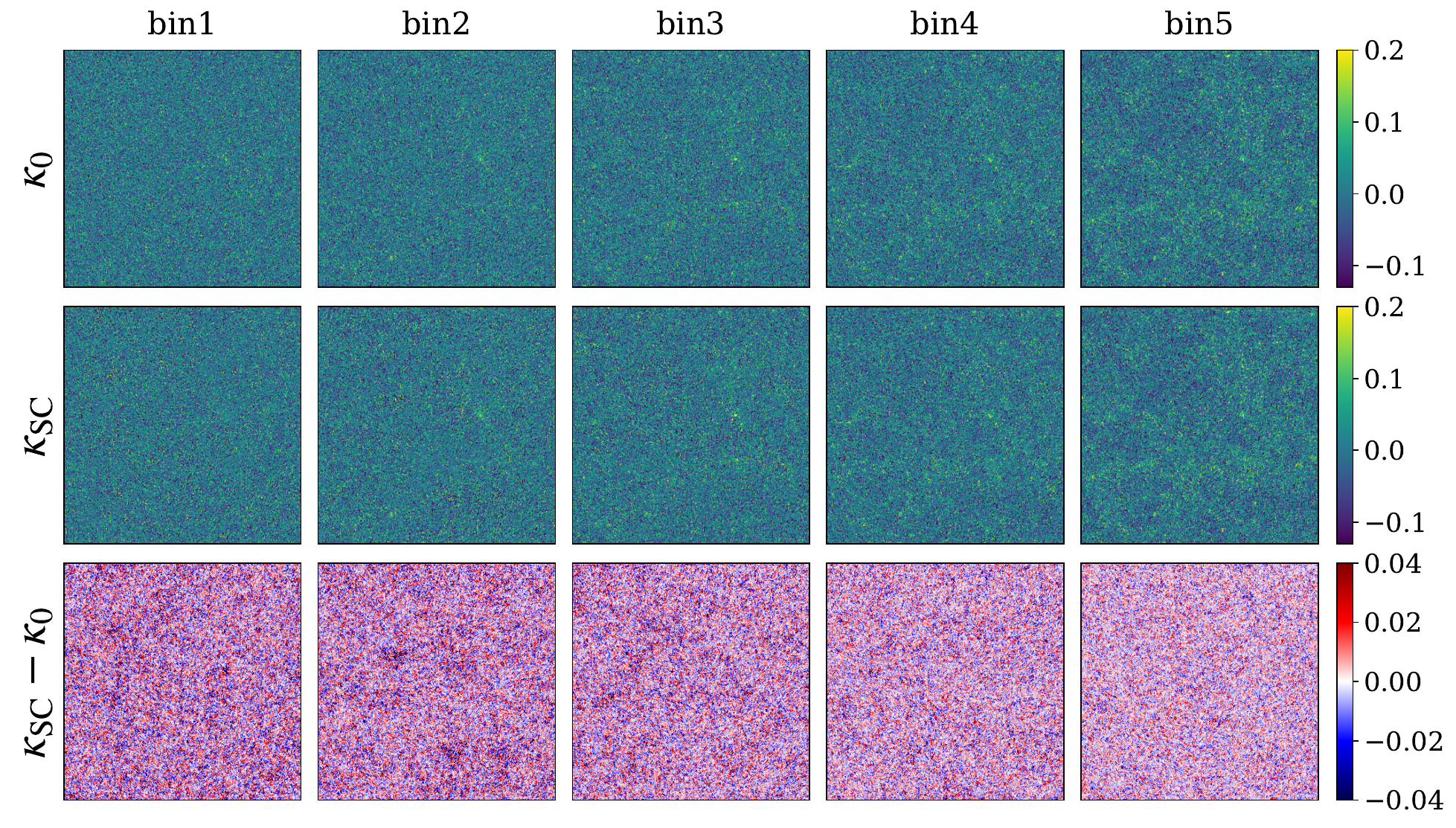}
    \caption{Same as Fig.~\ref{fig02:convergence_noiseless} but for noisy convergence maps.}
\label{fig03:convergence_noisy}
\end{figure*}

Fig.~\ref{fig02:convergence_noiseless} shows an example of noiseless convergence maps for the five tomographic bins at a given cosmology.
Source clustering has a larger impact at low redshift, introducing larger differences around clustered regions of high convergence. Although its impact decreases with redshift, it remains visible in all five tomographic bins.

The corresponding maps after adding shape noise are presented in Fig.~\ref{fig03:convergence_noisy}.
The spatial variations introduced by shape noise dominate over the differences caused by source clustering, making the latter difficult to identify by visual inspection.
Nevertheless, source clustering modulates the local noise standard deviation with source density, which can bias higher-order summary statistics.

%%%%%%%%%%%%%%%%%%%%%%%%%%%%%%%%%%%%%%%%%%%%%%%%%%%%%%%%%%%%%%
\section{Weak lensing analysis with summary statistics}
\label{sect:summary_statistics}
\begin{table*}[ht]
    \caption{Summary of weak lensing statistics used in this work, with their filtering methods, data vector lengths, and computational algorithms.}
    \centering
    \begin{tabular}{ccccc}
    \hline\hline
    Statistic & Abbreviation & Filter & Data vector length & Algorithm \\
    \hline
    Third-order moments & $\langle\kappa^3\rangle$ & Top-hat & 250\tablefootmark{a} & Central moment \\
    \hline
    Wavelet peak counts & $N_{\text{peak}}$ & Starlet wavelets & 300\tablefootmark{b} & \texttt{wl-stats-torch} \\
    Wavelet $\ell_1$-norm & $\ell_1$-norm & Starlet wavelets & 300\tablefootmark{b} & \texttt{wl-stats-torch} \\
    \hline
    \end{tabular}
    % \tablefoot{}
    \tablefoottext{a}{$(10\ \text{smoothing scales}) \times (25\ \text{bin combinations})$}
    \tablefoottext{b}{$(20\ \mathrm{S/N}\ \text{bins}) \times (3\ \text{wavelet scales}) \times (5\ \text{tomographic bins})$}
    \label{tab01:data_vectors}
\end{table*}

In this section, we describe the higher-order summary statistics considered in this work, namely third-order moments, wavelet peak counts and the wavelet $\ell_1$-norm of the convergence field.
We apply different filters to the convergence maps for each summary statistic.
We denote filters in real space by $F$ and in Fourier space by $\hat{F}$. We use $\kappa$ to denote the filtered convergence maps, with subscripts indicating filter parameters (e.g. the smoothing scale $\theta$ or wavelet scale $s$) and superscripts denoting tomographic bin indices.
The data vectors resulting from combining tomographic bins and filter parameters are summarised in Table~\ref{tab01:data_vectors}.

\subsection{Third-order moments}
\label{sect:moment3}
Third-order moments quantify the skewness of the convergence field.
\cite{Gatti24} measured the impact of source clustering in the measurements of the third-order moments. We quantify the bias this effect introduces in the cosmological constraints from third-order moment analyses.

We measure the central third-order moments of the convergence field after smoothing the convergence maps with a top-hat filter.
The Fourier transform of the top-hat filter is given by
\begin{equation}
    \hat{F}^{\mathrm{TH}}_\theta(\boldsymbol{k})
    =\frac{2J_1(k\theta)}{k\theta},
    \label{eq02:filter_tophat}
\end{equation}
where \(J_1\) is the Bessel function of the first kind of order one, \(k=|\boldsymbol{k}|\), and \(\theta\) is the smoothing scale.
The smoothed convergence map in tomographic bin \(i\) is then
\begin{equation}
    \kappa_\theta^i(\boldsymbol{x})
    =\left(\kappa^i * F_\theta^{\mathrm{TH}}\right)(\boldsymbol{x}).
    \label{eq03:filtered_tophat_convergence}
\end{equation}

We compute the third-order central moments for 25 tomographic bin
combinations of the form $(i,j,j)$, where $i,j\in\{1,\ldots,5\}$,
\begin{equation}
    \left<\kappa^3\right>^{i,j,j}(\theta)
    =
    \frac{1}{N_{\mathrm{pix}}}
    \sum_{p=1}^{N_{\mathrm{pix}}}
    \left(\kappa_\theta^i(p)-\overline{\kappa}_\theta^i\right)
    \left(\kappa_\theta^j(p)-\overline{\kappa}_\theta^j\right)^2,
    \label{eq04:stats_moment3}
\end{equation}
where $N_{\mathrm{pix}}$ is the number of pixels and
$\overline{\kappa}_\theta^i$ denotes the spatial mean of the smoothed
convergence map in tomographic bin $i$.

We construct the data vector using 10 logarithmically spaced smoothing scales, ranging from 3.44 arcmin (twice the pixel resolution) to 300 arcmin (half the patch size).
Combined with the 25 tomographic bin combinations, this yields a data vector of length 250 for each convergence map.

\subsection{Wavelet statistics}
\label{sect:wavelet_statistics}
Wavelet transforms provide a multiscale representation of a signal, enabling structures to be analysed at different spatial scales while preserving their spatial localisation \citep{Starck2007_wavelet}.
We use the Starlet wavelet transform to probe non-Gaussian features in the convergence field through two summary statistics: wavelet peak counts \citep{Pires2009_WPC, Ajani2020_WPC, 2025A&A...698A..25T} and the $\ell_1$-norm of the wavelet coefficients \citep{Ajani2021_l1, Sreekanth2024}.
We compute both statistics using the public \texttt{wl-stats-torch} package \citep{wl_stats_torch}, which provides GPU-accelerated implementations of wavelet-based summary statistics using \texttt{PyTorch}.

We use the Starlet wavelet transform to compute the wavelet peak counts and the wavelet $\ell_1$-norm.
The Starlet transform is constructed from the cubic $B_3$-spline scaling function $\phi_1$,
\begin{equation}
    \phi_1(x)=\frac{1}{12}\left(|x-2|^3\!-4|x-1|^3\!+6|x|^3\!-4|x+1|^3\!+|x+2|^3\right).
    \label{eq05:scaling_function}
\end{equation}
Given a reference scale \(\theta\), the two dimensional scaling function at dyadic scale \(s\) is defined as
\begin{equation}
    \phi_{s,\theta}(\boldsymbol{x})
    =
    \frac{1}{(2^s\theta)^2}
    \phi_1\!\left(\frac{x}{2^s\theta}\right)
    \phi_1\!\left(\frac{y}{2^s\theta}\right).
    \label{eq06:scaling_function_2d}
\end{equation}
The corresponding Starlet wavelet filter is the difference between two consecutive scaling functions,
\begin{equation}
    F_{s,\theta}^{\mathrm{SW}}(\boldsymbol{x})
    =
    \phi_{s,\theta}(\boldsymbol{x})
    -\phi_{s+1,\theta}(\boldsymbol{x}).
    \label{eq07:filter_starlet}
\end{equation}
Applying this filter to the convergence map yields the wavelet coefficient map,
\begin{equation}
    \kappa_{s,\theta}^i(\boldsymbol{x})
    =
    \left(\kappa^i * F_{s,\theta}^{\mathrm{SW}}\right)(\boldsymbol{x}).
    \label{eq08:filtered_starlet_convergence}
\end{equation}

The original convergence map can be reconstructed from the wavelet coefficient maps and the coarse resolution component $c_S^i$ as
\begin{equation}
    \kappa^i(\boldsymbol{x})
    =
    \sum_{s=1}^{s_{\max}}
    \kappa_{s,\theta}^i(\boldsymbol{x})+c_S^i(\boldsymbol{x}).
    \label{eq09:filtered_starlet_convergence_full}
\end{equation}
The two wavelet coefficient maps used in this work have nominal
angular scales of $3.44'$ and $6.88'$, while the coarse resolution
component corresponds to the largest scale.
For simplicity, we treat the coarse resolution component as the wavelet map at the largest scale.
Thus, of the three scales used in this work, the first two correspond to wavelet coefficient maps $\kappa_{s,\theta}^i(\boldsymbol{x})$ and the third to the coarse resolution component $c_S^i(\boldsymbol{x})$.
In the following, we use these maps to define the wavelet peak counts and the $\ell_1$-norm.

\subsubsection{Wavelet peak counts}
\label{sect:wavelet_peak}
Wavelet peak counts quantify the abundance of local maxima in the convergence field at different spatial scales.
In each Starlet wavelet map $\kappa_{s,\theta}^i(\boldsymbol{x})$, we identify local maxima by selecting pixels whose values are greater than those of their eight neighbouring pixels.
The noise standard deviation at each wavelet scale is computed by convolving the pixel-wise noise variance with the squared wavelet impulse response at that scale,
\begin{equation}
    \left[\sigma_{N,s}^i(\boldsymbol{x})\right]^2
    =
    \left[
        \left(\sigma_{\mathrm{pix}}^i\right)^2
        * \left(F_{s,\theta}^{\mathrm{SW}}\right)^2
    \right](\boldsymbol{x}),
    \label{eq10:wavelet_noise}
\end{equation}
where $\sigma_{\mathrm{pix}}^i(\boldsymbol{x})$ is the pixel-wise shape noise standard deviation in tomographic bin $i$.
The signal-to-noise ratio (S/N) at each peak, $\nu_s^i(\boldsymbol{x})$, is then defined as the ratio of the wavelet coefficient to the corresponding noise standard deviation,
\begin{equation}
    \nu_s^i(\boldsymbol{x})
    =
    \frac{\kappa_{s,\theta}^i(\boldsymbol{x})}
    {\sigma_{N,s}^i(\boldsymbol{x})}.
    \label{eq11:wavelet_snr}
\end{equation}
When source clustering is included, it also affects the standard deviation of the pixel noise via the spatially varying galaxy number density, thereby entering the S/N normalisation itself.

We construct a histogram of the peak S/N values using 20 bins spanning the range $[-5,5]$, which covers the range where most of the signal is concentrated.
Repeating this procedure across the three scales and five tomographic bins yields a 300-element data vector for each convergence map.

\subsubsection{Wavelet \texorpdfstring{$\ell_1$}{l1}-norm}
\label{sect:wavelet_l1}
The wavelet $\ell_1$-norm quantifies the sum of the absolute values of noise-normalised wavelet coefficients within S/N bins.
Using the S/N values $\nu_s^i(\boldsymbol{x})$ defined in Eq.~(\ref{eq11:wavelet_snr}), we define the set of pixels belonging to the $b$-th S/N bin as
\begin{equation}
    S_{s,b}^i
    =
    \left\{
        \boldsymbol{x} \ \middle| \;
        B_b < \nu_s^i(\boldsymbol{x}) < B_{b+1}
    \right\},
    \label{eq12:wavelet_l1_bin}
\end{equation}
where $B_b$ and $B_{b+1}$ are the lower and upper boundaries of the S/N bin, respectively.
The wavelet $\ell_1$-norm is then computed as
\begin{equation}
    \left[\ell_1\right]_{s,b}^i
    =
    \sum_{\boldsymbol{x}\in S_{s,b}^i}
    \left|\nu_s^i(\boldsymbol{x})\right|.
    \label{eq13:wavelet_l1}
\end{equation}

Using the same S/N binning as for wavelet peak counts, we compute the $\ell_1$-norm in 20 bins spanning the range $[-5,5]$.
Combining the three scales and five tomographic bins yields a data vector of length 300 for each convergence map.

%%%%%%%%%%%%%%%%%%%%%%%%%%%%%%%%%%%%%%%%%%%%%%%%%%%%%%%%%%%%%%
\section{Methodology}
\label{sect:methodology}
\subsection{Simulation-based inference}
\label{sect:sbi}
We use simulation-based inference (SBI) to constrain cosmological parameters using higher-order weak lensing summary statistics.
SBI enables posterior estimation directly from simulations without requiring an explicit likelihood function.
We perform inference using the \texttt{LtU-ILI} framework \citep{Ho2024}, which uses simulated cosmological parameters and their corresponding summary statistics to estimate the posterior distribution.

To quantify the bias induced by source clustering in cosmological inference, we use the Gower Street simulations to build training, validation, and test sets.
We consider 599 cosmologies, each with 14 patches at different positions, both with and without source clustering.
The training and validation sets do not include source clustering, and we assign 80\% of the cosmologies (479) to the training set and the remaining 20\% (120) to the validation set.
The test set consists of the source clustering versions of the same 120 validation cosmologies.
We compare the resulting constraints with those obtained without source clustering to quantify the bias introduced by neglecting this effect in the modelling.
In addition, we perform a complementary analysis using the same splitting procedure, but including source clustering in both the training and test sets.
Comparing the results with and without source clustering in the training set allows us to assess the impact of model misspecification.

We use neural posterior estimation to train a neural density estimator that approximates the posterior distribution of the cosmological parameters given weak-lensing summary statistics.
Each training run uses only one of the considered summary statistics.

\begin{table}[ht!]
    \caption{Prior distributions for cosmological parameters.}
    \centering
    \begin{tabular}{cc}
    \hline\hline
    Parameter & Prior \\
    \hline
    $\Omega_\mathrm{m}$ & $\mathcal{U}(0.15, 0.52)$ \\
    $S_8$ & $\mathcal{U}(0.5, 1.0)$ \\
    $w$ & $\mathcal{U}(-1, 1/3)$ \\
    $h$ & $\mathcal{N}(0.7022, 0.0245)$ \\
    $\Omega_\mathrm{b}h^2$ & $\mathcal{N}(0.02237, 0.00015)$ \\
    $n_s$ & $\mathcal{N}(0.9649, 0.0063)$ \\
    $\log(m_\nu)$ & $\mathcal{U}(\log(0.06), \log(0.14))$ \\
    \hline
    \end{tabular}
    \label{tab02:sbi_prior}
\end{table}

Table~\ref{tab02:sbi_prior} shows the prior distributions for the neural posterior estimator in the SBI analyses.
We adopt the same prior distributions as \cite{Gatti2024_sbi}, based on the cosmological parameter space spanned by the Gower Street simulations \citep{Jeffrey2025_GowerSt}.
While the Gower Street simulations are parameterized in terms of $\sigma_8$, we use $S_8 = \sigma_8\sqrt{\,\Omega_{\mathrm{m}}/0.3\,}$ instead as an inference parameter.

\subsection{Pipeline validation}
\label{sect:sbi_validation}
We validate the SBI posterior on the validation set, without source clustering, to assess whether the training procedure reliably recovers the underlying cosmological parameters.
We examine posterior calibration through marginal coverage tests for individual cosmological parameters and the Tests of Accuracy with Random Points (TARP; \citealt{Lemos2023}) for the joint posterior.
The coverage tests assess whether the true parameter values fall within the corresponding marginal credible intervals with the expected frequency, while TARP assesses the calibration of the joint posterior.
The validation results are presented in Appendix~\ref{sect:appx_validation}.

We further examine the robustness of the inference results to the training stochasticity by repeating the procedure with 100 different random data splits.
For each split, the main analysis uses 479 cosmologies without source clustering for training.
The remaining 120 cosmologies are used for validation without source clustering and for testing with source clustering.
In a complementary analysis, we include source clustering in both the training and test data to compare the figure of merit of each statistic.
We find that the results remain consistent across the different splits, indicating that the inferred constraints are robust to the choice of cosmological partition.

%%%%%%%%%%%%%%%%%%%%%%%%%%%%%%%%%%%%%%%%%%%%%%%%%%%%%%%%%%%%%%
\section{Results}
\label{sect:results}
In this section, we first discuss the impact of source clustering on higher-order summary statistics across different tomographic bins and scales. We then quantify the bias this introduces to the inferred cosmological parameters.

\subsection{Impact of source clustering in summary statistics}
\label{sect:results_summary_statistics}
\begin{figure*}[ht!]
\centering
\includegraphics[width=170mm]{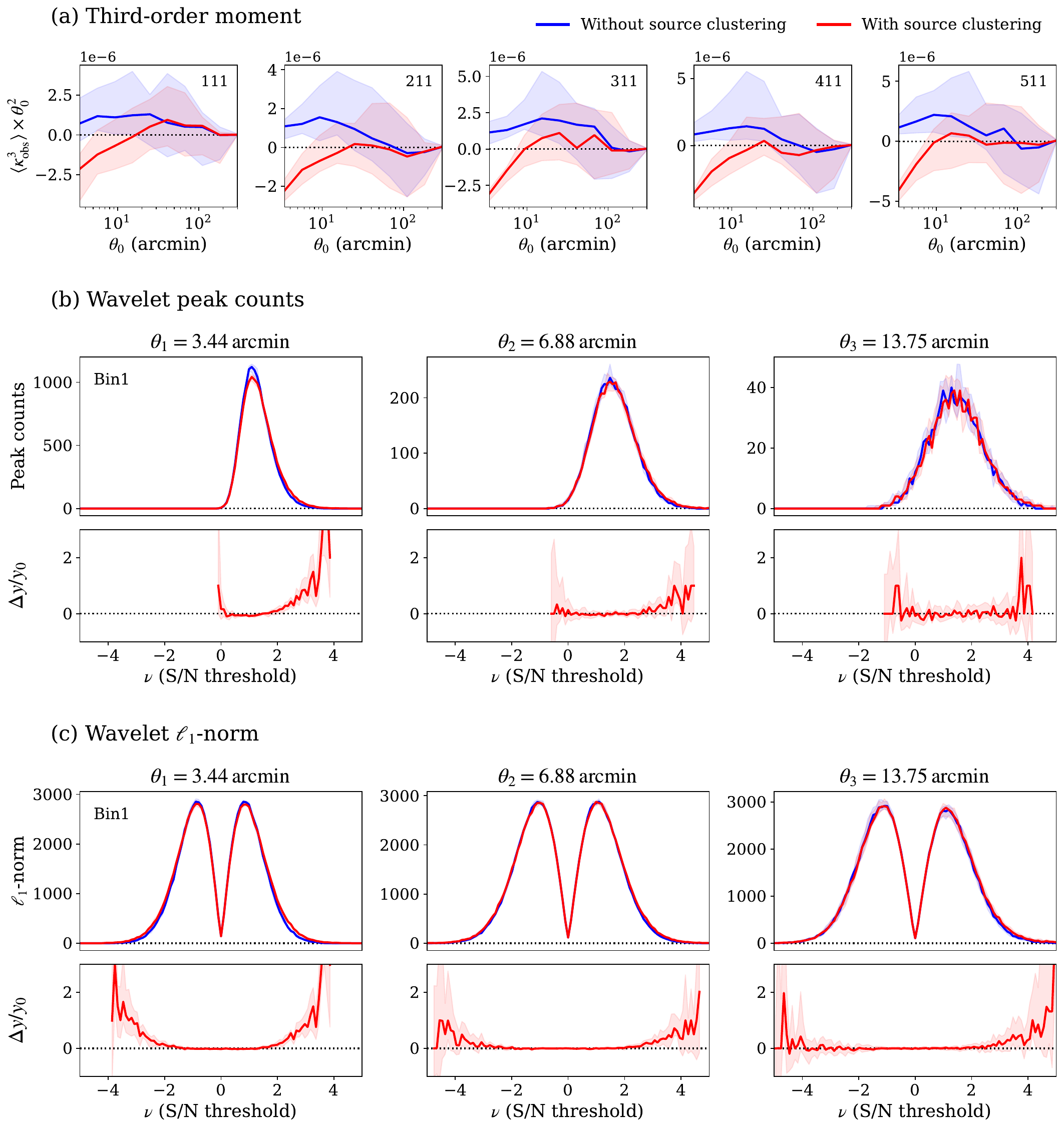}
    \caption{Data vectors for selected tomographic bins and scales:
    (a) third-order moments for the auto-correlation of tomographic bin 1, $(111)$, and its cross-correlations, $(211)$, $(311)$, $(411)$, and $(511)$;
    (b, c) wavelet peak counts and $\ell_1$-norm for tomographic bin 1, respectively.
    Blue and red indicate results without and with source clustering, respectively.}
\label{fig04:data_vectors_subset}
\end{figure*}

We report the median and 68\% interval across the 14 patches for a fiducial cosmology.\footnote{$\Omega_\mathrm{m}=0.3088,\ \sigma_8=0.7550,\ n_s=0.9623,\ h=0.7157,\ \Omega_\mathrm{b}h^2=0.0224,\ w=-0.6384,\ \text{and}\ m_\nu=0.0633$}
Figure~\ref{fig04:data_vectors_subset} shows tomographic bins of the three summary statistics, highlighting the regions where the impact of source clustering is most pronounced.
For the third-order moments, the largest deviations are observed in low-redshift tomographic bin combinations and at small smoothing scales, which is consistent with previous findings \citep{Gatti24}.
These differences are largely driven by the correlation between the convergence field and shape noise, arising from the spatial modulation of the noise by the source density.
For the wavelet peak counts and $\ell_1$-norm, the impact of source clustering is largest in the first tomographic bin and at the smallest wavelet scale.
The largest deviations occur in the tails of the distributions: at $\nu \gtrsim 2$ for peak counts, and at $\nu \lesssim -2$ and $\nu \gtrsim 2$ for the $\ell_1$-norm.
Regions with large absolute S/N values are more strongly affected by source clustering, which modulates the shape noise according to the local source density.

\subsection{Impact of source clustering in cosmology constraints}
\label{sect:results_cosmology_constraints}
We first compare the posterior distributions of cosmological parameters from two test datasets generated with the same true cosmology and the same sky patch, with and without source clustering. The shape-noise realisations are generated with matching random seeds for the maps with and without source clustering. 
The SBI pipeline is trained using 100 different random data splits to examine the sensitivity of the inference results to data partitioning.
Using the posterior trained on one selected split, we analyse the inference results for each test patch to quantify the impact of source clustering.
The results for each test patch are referred to as patch-level results, which account for the dependency of the source clustering bias on spatial variations and cosmic variance.
We then calculate the median of the patch-level results over the test patches within each data split, and examine the distribution of these medians across the 100 splits.
This distribution, referred to as the split-level results, indicates the impact of training stochasticity on the bias but does not reflect statistical uncertainty. The uncertainty on the bias is shown in the patch-level measurements.
We focus hereafter on $\Omega_{\mathrm{m}}$, $S_8$, and $w$, for which the SBI pipeline yields informative constraints.\footnote{The posteriors of $n_s$, $h$, $\Omega_\mathrm{b}h^2$, and $m_\nu$ are prior-dominated and are therefore not discussed further.}

\begin{figure}%[ht!]
\centering
\includegraphics[width=0.845\hsize]{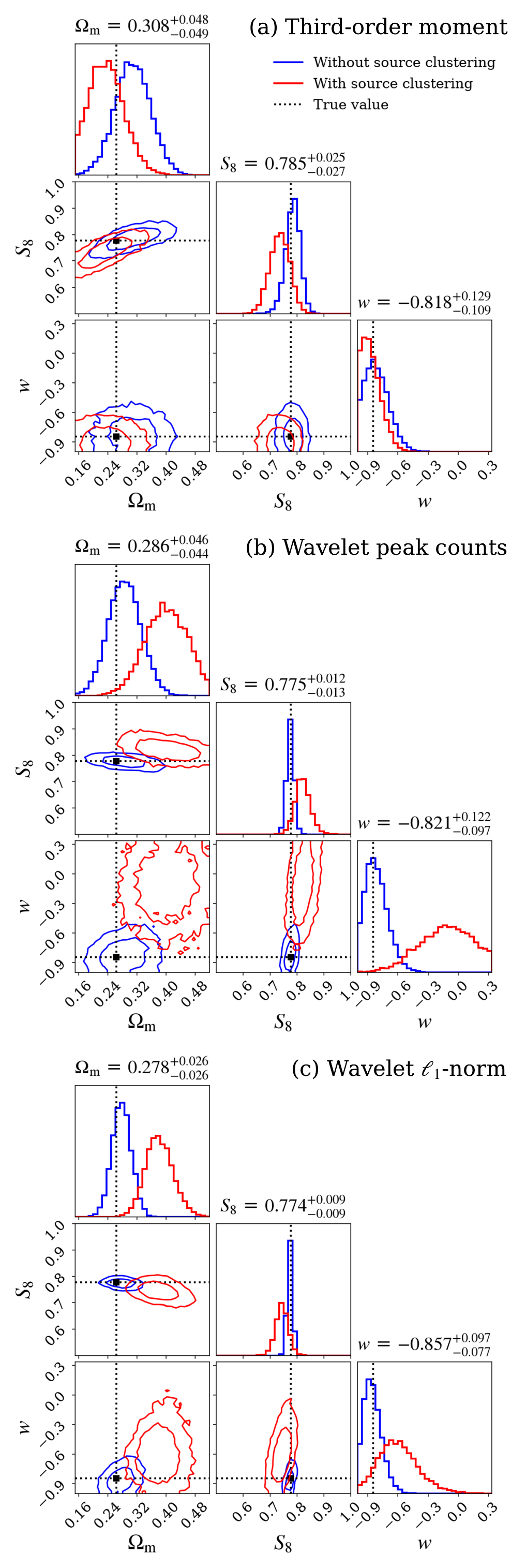}
    \caption{Posterior distributions of $\Omega_{\mathrm{m}}$, $S_8$, and $w$ for (a) third-order moments, (b) wavelet peak counts, and (c) wavelet $\ell_1$-norm. Blue and red contours show results for simulated data without and with source clustering, respectively. In all cases, the data model does not include source clustering. Black dots indicate the true cosmological parameters.}
\label{fig05:sbi_posterior_bias}
\end{figure}

Fig.~\ref{fig05:sbi_posterior_bias} shows the posterior distributions of $\Omega_\mathrm{m}$, $S_8$, and $w$ from the analyses of the three summary statistics: third-order moments, wavelet peak counts, and the wavelet $\ell_1$-norm.
We find that source clustering significantly biases the cosmological constraints obtained from all three summary statistics if this effect is not accounted for in the modelling.
The magnitude and direction of the bias depend on the summary statistic.
Wavelet statistics exhibit larger biases than third-order moments in $\Omega_\mathrm{m}$ and $w$, while the biases in $S_8$ are in opposite directions. 
In addition, the posterior contours of both wavelet statistics broaden when the target data contain source clustering effects, indicating a degradation in parameter constraints due to mismodeling of the source clustering.
\cite{Schneider2020} also reported a degradation in weak lensing constraints due to model misspecification.
We further quantify this degradation using a metric of constraining power.

To account for cosmic variance in these results, we introduce an uncertainty-normalised bias estimator based on the difference between the posterior medians obtained with and without source clustering.
For each test patch, we define the estimator as
\begin{equation}
    B_{\theta}
    =
    \frac{
    \theta_{\mathrm{SC}}^{\mathrm{med}}
    -
    \theta_{0}^{\mathrm{med}}
    }{
    \sqrt{
    \sigma_{\mathrm{SC}}^2
    +
    \sigma_{0}^2
    }
    },
    \label{eq14:bias_estimator}
\end{equation}
where $\theta^{\mathrm{med}}$ denotes the posterior median and $\sigma$ represents the posterior standard deviation.
The subscripts $\mathrm{SC}$ and $0$ indicate the analyses with and without source clustering, respectively.

\begin{figure}[ht!]
\centering
\includegraphics[width=85mm]{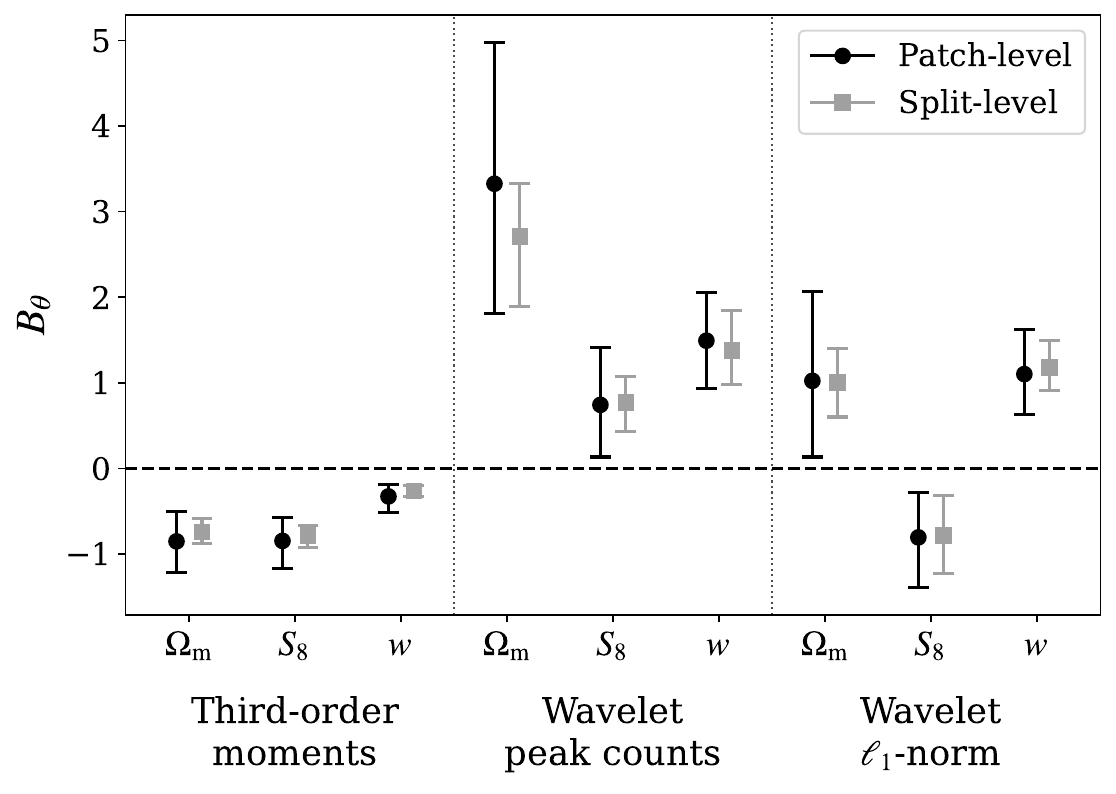}
    \caption{Source clustering bias for $\Omega_{\mathrm{m}}$, $S_8$, and $w$ across the three summary statistics. Circles and squares indicate patch-level and split-level results, respectively.}
\label{fig06:bias_sc}
\end{figure}

\begin{table}
\caption{Source clustering bias in cosmological parameter constraints. Values indicate the median bias, with uncertainties corresponding to the 16th--84th percentile range.}
\centering
\renewcommand{\arraystretch}{1.5}
\begin{tabular}{cccc}
\hline\hline
\multirow{2}{*}{\shortstack[c]{Summary\\statistic}} & \multirow{2}{*}{Parameter} & \multicolumn{2}{c}{Bias ($B_\theta$)} \\
 & & Patch-level & Split-level \\
\hline
\multirow{3}{*}{\shortstack[c]{Third-order\\moments}} & $\Omega_{\mathrm{m}}$ & ${-0.853}^{+0.352}_{-0.363}$ & ${-0.743}^{+0.154}_{-0.139}$ \\
 & $S_8$ & ${-0.847}^{+0.278}_{-0.321}$ & ${-0.781}^{+0.113}_{-0.142}$ \\
 & $w$ & ${-0.328}^{+0.135}_{-0.185}$ & ${-0.262}^{+0.060}_{-0.062}$ \\
\hline
\multirow{3}{*}{\shortstack[c]{Wavelet\\peak counts}} & $\Omega_{\mathrm{m}}$ & ${3.326}^{+1.653}_{-1.515}$ & ${2.710}^{+0.620}_{-0.818}$ \\
 & $S_8$ & ${0.743}^{+0.669}_{-0.609}$ & ${0.769}^{+0.301}_{-0.339}$ \\
 & $w$ & ${1.492}^{+0.561}_{-0.561}$ & ${1.377}^{+0.467}_{-0.393}$ \\
\hline
\multirow{3}{*}{\shortstack[c]{Wavelet\\$\ell_1$-norm}} & $\Omega_{\mathrm{m}}$ & ${1.023}^{+1.041}_{-0.890}$ & ${1.003}^{+0.399}_{-0.403}$ \\
 & $S_8$ & ${-0.806}^{+0.520}_{-0.588}$ & ${-0.784}^{+0.472}_{-0.443}$ \\
 & $w$ & ${1.102}^{+0.520}_{-0.472}$ & ${1.180}^{+0.316}_{-0.271}$ \\
\hline
\end{tabular}
\label{tab03:bias_sc}
\end{table}

Fig.~\ref{fig06:bias_sc} and Table~\ref{tab03:bias_sc} summarise the median bias and its 16th--84th percentile range at both the patch and split levels.
At the patch-level, we derive these percentiles from the bias distribution across all 1,680 targets (120 cosmologies $\times$ 14 patches) within a given split.
For split-level results, we use the distribution of 100 median bias values, each summarising all 1,680 targets in a different split.
The patch-level results exhibit a larger spread than the split-level results, indicating the sensitivity of the bias to cosmic variance and the robustness of the results to the training set.
The sensitivity to cosmic variance and spatial variations is more pronounced for the wavelet statistics than for the third-order moments.
The third-order moments consistently bias the cosmological parameters towards lower values, with similar posterior shifts for $\Omega_{\mathrm{m}}$ and $S_8$ and a smaller bias in $w$. 
Wavelet peak counts bias the results towards higher values, with the largest bias found for $\Omega_{\mathrm{m}}$, reaching $3.326$.
The wavelet $\ell_1$-norm also biases  $\Omega_{\mathrm{m}}$ and $w$ towards larger values, but underestimates $S_8$. 

\begin{figure*}[ht!]
\centering
\includegraphics[width=170mm]{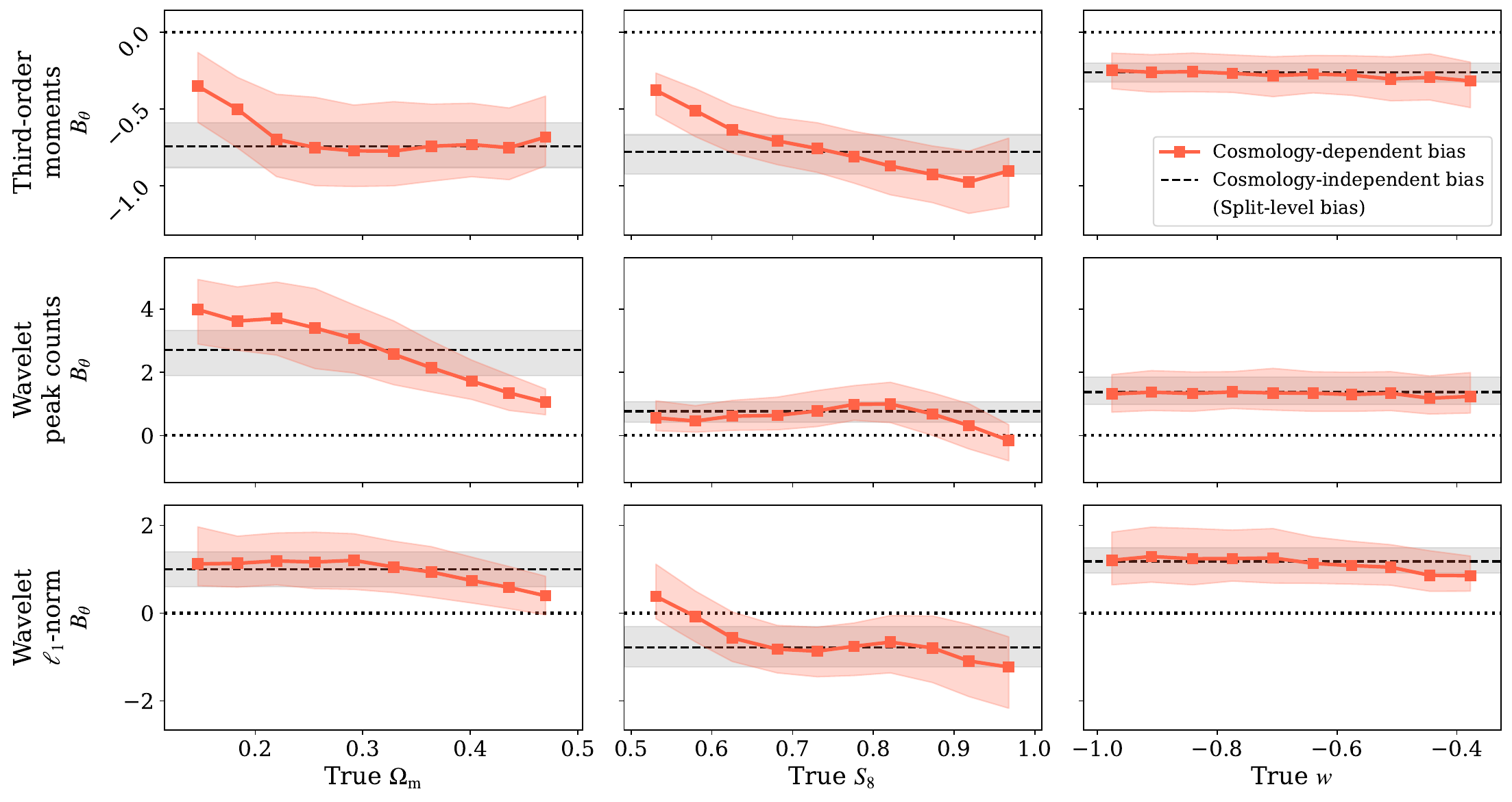}
    \caption{Dependency of the source clustering bias on the underlying true cosmological parameters. The orange region shows the binned bias for different underlying cosmologies, while the grey region shows the cosmology-independent bias (cosmic variance computed from the split-level analyses), as presented in Fig.~\ref{fig06:bias_sc}.}
\label{fig07:bias_sc_cosmo}
\end{figure*}

The bias estimator is computed by averaging over different cosmologies.
That might obscure potential sensitivity of the source clustering bias to the underlying true cosmology, leading to an underestimation of the bias if it changes in magnitude or direction.
To investigate this cosmology dependence, we divide the datasets into 10 bins based on their true parameter values and compute the distribution of median biases for each bin.
Fig.~\ref{fig07:bias_sc_cosmo} shows the dependence of the bias on the underlying true cosmological parameters.
For most summary statistics, the source clustering bias is independent of the true cosmology.
However, third-order moments show a negative correlation between the bias and the true value of $S_8$, with a Pearson correlation coefficient of $-0.586$.
As a result, the bias magnitude increases with the true value of $S_8$.
The wavelet peak counts also show a moderate negative correlation with the true value of $\Omega_{\mathrm{m}}$, with a Pearson coefficient of $-0.575$, indicating larger bias magnitudes at lower $\Omega_{\mathrm{m}}$ values.
The wavelet $\ell_1$-norm shows a negligible bias magnitude at low $S_8$ values, values close to the split-level bias at intermediate $S_8$, and a slight increase towards higher $S_8$ values.

As the greater sensitivity to source clustering might be connected to greater constraining power, we also calculate the figure of merit (FoM) for each summary statistic.
For each test target, we draw 1,000 samples from the joint posterior distribution estimated by the trained neural posterior estimator.
We then estimate the covariance matrix of the three cosmological parameters, $\Omega_{\mathrm{m}}$, $S_8$, and $w$, from these posterior samples.
The FoM is defined as
\begin{equation}
    \mathrm{FoM}(\boldsymbol{\theta})
    =
    \det\left(\mathbf{C}_{\boldsymbol{\theta}}^{-1}\right)^{1/n},
    \label{eq15:fom}
\end{equation}
where $\mathbf{C}_{\boldsymbol{\theta}}$ is the covariance matrix of the joint posterior distribution of the cosmological parameters under consideration, and $n$ is the dimensionality of the parameter space.
A larger FoM indicates tighter constraints on the cosmological parameters, corresponding to a smaller posterior volume.

\begin{figure}[ht!]
\centering
\includegraphics[width=85mm]{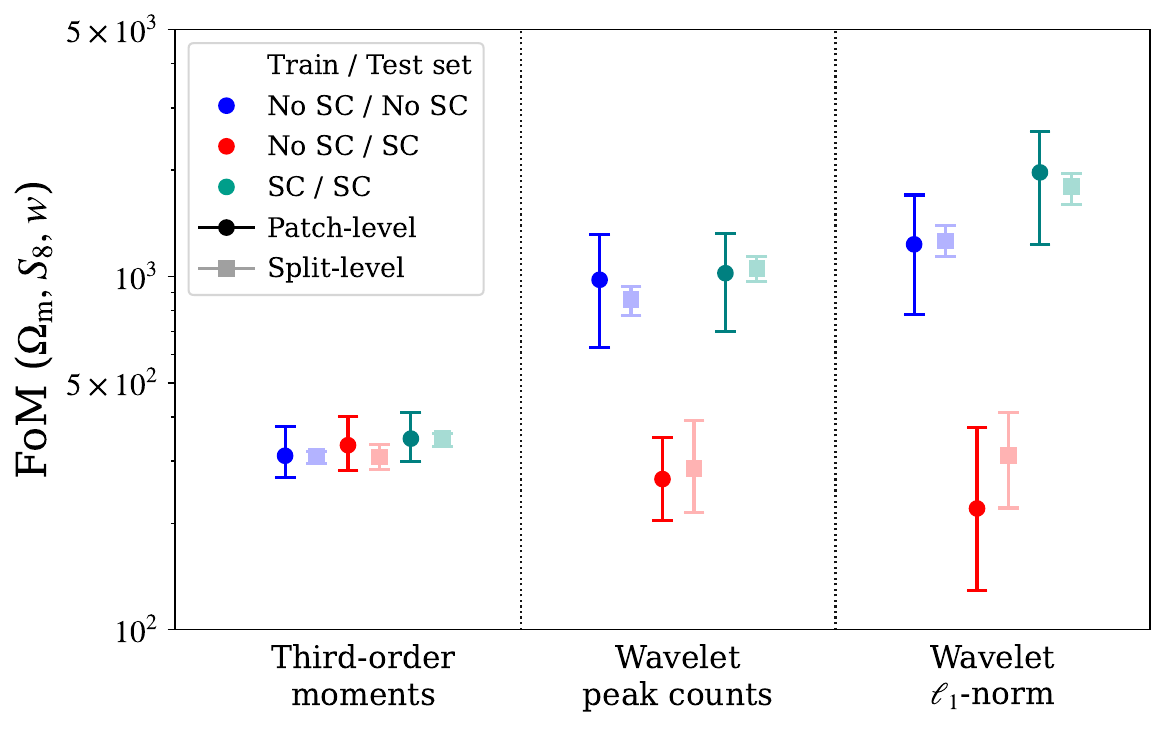}
    \caption{Figure of merit for the three summary statistics, considering the $\Omega_\mathrm{m}$, $S_8$, and $w$ parameter combination. Blue, red, and teal correspond to No SC / No SC, No SC / SC, and SC / SC, respectively, where the first and second entries refer to the training and test sets.}
\label{fig08:fom_sc}
\end{figure}

Fig.~\ref{fig08:fom_sc} summarises the median FoM and its 16th--84th percentile range at both the patch and split levels, considering the joint constraints on $(\Omega_{\mathrm{m}}, S_8, w)$.
Without source clustering, the wavelet statistics yield higher FoM values than the third-order moments.
The inclusion of source clustering has negligible impact on the FoM obtained from the third-order moments.
In contrast, the FoM decreases by 72.7\% for the wavelet peak counts and 82.1\% for the wavelet $\ell_1$-norm when source clustering is included in the target data but neglected in the training model.
These results show that mismodeling of the data by neglecting source clustering in the training model degrades the constraining power and inflates the contours.
This leads to a considerable loss of cosmological information compared to the case without source clustering.
However, when source clustering is consistently included in both the training and test sets, the FoM recovers and even exhibits a slight increase.
This improvement is due to source clustering also carrying cosmological information.

%%%%%%%%%%%%%%%%%%%%%%%%%%%%%%%%%%%%%%%%%%%%%%%%%%%%%%%%%%%%%%
\section{Conclusions}
\label{sect:conclusions}
In this work, we have investigated the impact of source clustering on weak lensing higher-order statistics using simulation-based inference.
We considered three summary statistics: third-order moments, wavelet peak counts, and the wavelet $\ell_1$-norm. Our main findings are summarised as follows.
\begin{itemize}
    \item{
    Source clustering affects all three summary statistics, and, when not accounted for in the modelling, it induces biases in the inferred cosmological parameters of up to $3\sigma$.
    The direction of the bias depends on the summary statistic: third-order moments generally shift the parameters towards lower values, whereas wavelet peak counts shift them towards higher values.
    The largest bias is found for $\Omega_{\mathrm{m}}$ from wavelet peak counts, reaching $3.326$.
    }
    \item{
    The impact of source clustering can also depend on the true cosmology.
    For most summary statistics and parameters, the bias is cosmology-independent.
    However, correlations between the bias and the true cosmological parameter values are observed for specific statistic-parameter combinations, highlighting the importance of accounting for cosmological dependence when assessing the impact of source clustering.
    } 
    \item{
    Mismodeling source clustering in the training model degrades the constraining power of weak-lensing summary statistics.
    While the FoM from third-order moments remains relatively stable, it decreases by 72.7\% for wavelet peak counts and 82.1\% for the wavelet $\ell_1$-norm when source clustering is neglected in the training model.
    }    
\end{itemize}

These findings highlight the need to account for source clustering when extracting cosmological information from weak-lensing observations, particularly at low redshifts and on small scales. 
Our additional scale cut analysis (Appendix~\ref{sect:appx_cut}) further demonstrates that appropriate scale selection can help mitigate its impact.
This will be important for the robust exploitation of Euclid weak-lensing data.
In a follow-up, we will include intrinsic alignments and their corresponding nuisance parameters and investigate how this affects the source clustering bias.     

%%%%%%%%%%%%%%%%%%%%%%%%%%%%%%%%%%%%%%%%%%%%%%%%%%%%%%%%%%%%%%
\begin{acknowledgements}
The authors thank Lina Castiblanco, Joachim Harnois-Deraps, Marco Gatti, Nicolas Martinet, Cora Uhlemann and Simone Vinciguerra for useful discussions. S.-g. H. is supported by the Bourse France Excellence. This work was supported by the OCAPi project, funded by the European Research Council (ERC) under the Horizon 2020 research and innovation program (grant agreement number 101220619, PI: Natalia Porqueres). JL is supported by the TITAN ERA Chair project 
(contract no. 101086741) within the Horizon Europe Framework Programme of the
European Commission and  the French Agence Nationale de la Recherche (ANR-22-
CE31-0014-01 TOSCA).
\end{acknowledgements}

%%%%%%%%%%%%%%%%%%%%%%%%%%%%%%%%%%%%%%%%%%%%%%%%%%%%%%%%%%%%%%
% WARNING
% Please note that we have included the references below in
% order to compile the document, but we ask you to:
%
% - use BibTeX with the regular commands:
%   \bibliographystyle{aa} % style aa.bst
%   \bibliography{Yourfile} % your references Yourfile.bib
\bibliographystyle{aa.bst}
\bibliography{biblio.bib}
% - join the .bib files when you upload your source files
%%%%%%%%%%%%%%%%%%%%%%%%%%%%%%%%%%%%%%%%%%%%%%%%%%%%%%%%%%%%%%

% \begin{thebibliography}{}
%   \bibitem[Baker(1966)]{baker} Baker, N. 1966,
%       in Stellar Evolution,
%       ed.\ R. F. Stein,\& A. G. W. Cameron
%       (Plenum, New York) 333
%    \bibitem[Balluch(1988)]{balluch} Balluch, M. 1988,
%       A\&A, 200, 58
%    \bibitem[Cox(1980)]{cox} Cox, J. P. 1980,
%       Theory of Stellar Pulsation
%       (Princeton University Press, Princeton) 165
%    \bibitem[Cox(1969)]{cox69} Cox, A. N.,\& Stewart, J. N. 1969,
%       Academia Nauk, Scientific Information 15, 1
%    \bibitem[Mizuno(1980)]{mizuno} Mizuno H. 1980,
%       Prog. Theor. Phys., 64, 544
%    \bibitem[Tscharnuter(1987)]{tscharnuter} Tscharnuter W. M. 1987,
%       A\&A, 188, 55
%    \bibitem[Terlevich(1992)]{terlevich} Terlevich, R. 1992, in ASP Conf. Ser. 31,
%       Relationships between Active Galactic Nuclei and Starburst Galaxies,
%       ed. A. V. Filippenko, 13
%    \bibitem[Yorke(1980a)]{yorke80a} Yorke, H. W. 1980a,
%       A\&A, 86, 286
%    \bibitem[Zheng(1997)]{zheng} Zheng, W., Davidsen, A. F., Tytler, D. \& Kriss, G. A.
%       1997, preprint
% \end{thebibliography}

%%%%%%%%%%%%%%%%%%%%%%%%%%%%%%%%%%%%%%%%%%%%%%%%%%%%%%%%%%%%%%%
% Appendices must be placed after   \end{thebibliography}
% They will be placed automatically on a new page.
%%%%%%%%%%%%%%%%%%%%%%%%%%%%%%%%%%%%%%%%%%%%%%%%%%%%%%%%%%%%%%%
\begin{appendix}
%%%%%%%%%%%%%%%%%%%%%%%%%%%%%%%%%%%%%%%%%%%%%%%%%%%%%%%%%%%%%%%
% In the PDF output, floats should be placed
% under their own appendix, not before the title, nor after the
% title of the next appendix.

% In short appendices, onecolumn floats (\figure*
% or \table*) will generate a blank page.
% To prevent this behaviour, a few examples are provided here. 

% In case you have a lot of floating objects for little text and the 
% LaTeX engine moves the floats away from their context, the command
% \FloatBarrier of the “placeins” package will empty the
% float buffer and place all stored floats in the continuity.

% If you still encounter problems with wide floats placement,
% just use the onecolumn environment throughout the appendices.
%%%%%%%%%%%%%%%%%%%%%%%%%%%%%%%%%%%%%%%%%%%%%%%%%%%%%%%%%%%%%%%

%____________________________________________________________
%       Wide floats at the start of an appendix: first method
%-------------------------------------------------------------
% To prevent a blank page after the start of an appendix:
% - Switch to one \onecolumn first
% - Declare the section title
% - Declare the onecolumn float with the parameter [ht!]
% - Revert to \twocolumn at the end of the section
% \onecolumn
%%%%%%%%%%%%%%%%%%%%%%%%%%%%%%%%%%%%%%%%%%%%%%%%%%%%%%%%%%%%%%
\section{Posterior validation}
\label{sect:appx_validation}

\begin{figure*}[ht!]
\centering
\includegraphics[width=170mm]{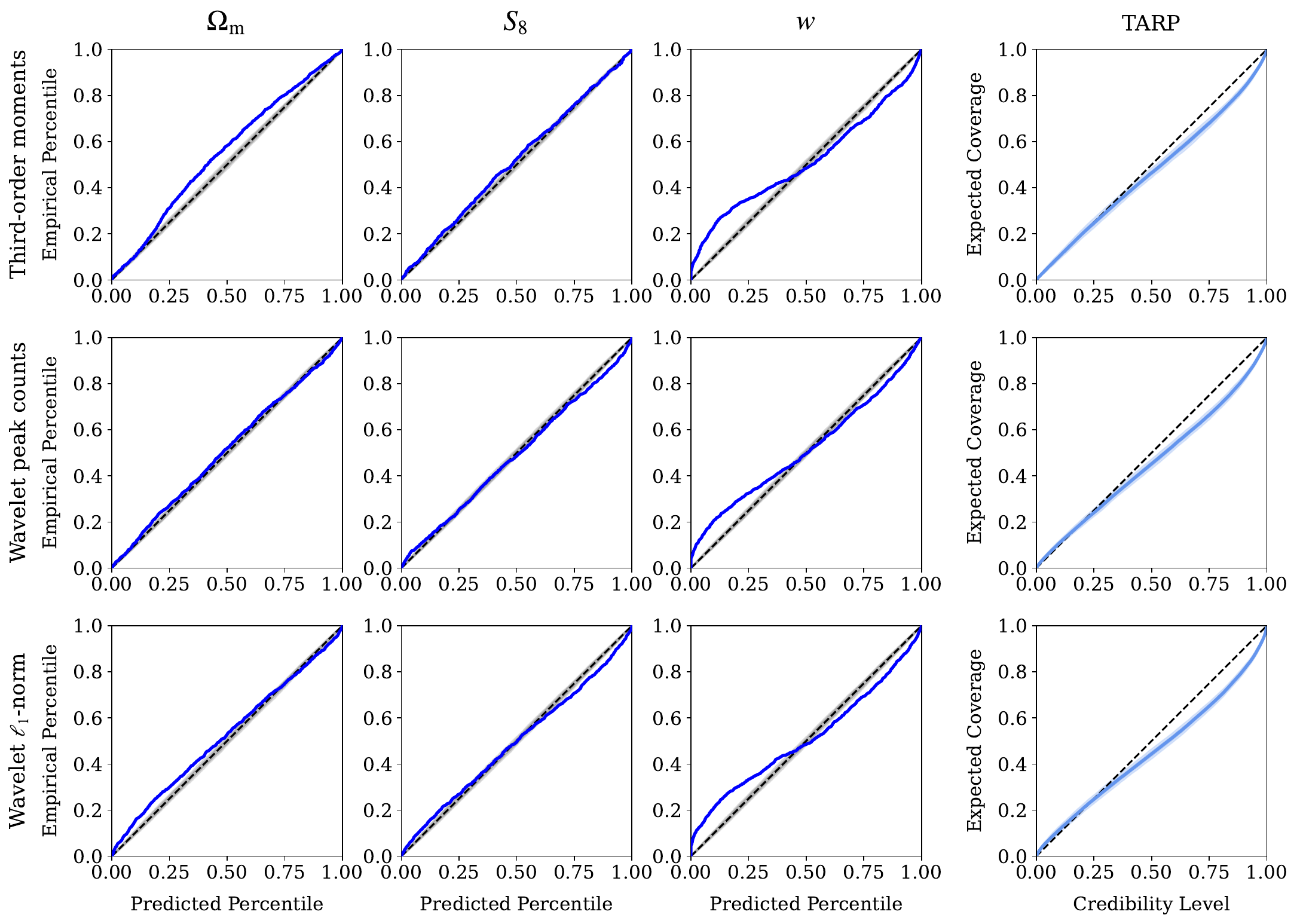}
    \caption{Posterior calibration results from the marginal coverage tests and TARP for the three summary statistics.}
\label{figa01:sbi_validation}
\end{figure*}

We present the posterior calibration results obtained from the validation set without source clustering in Fig.~\ref{figa01:sbi_validation}.
Overall, the marginal coverage tests show good calibration for $\Omega_\mathrm{m}$ and particularly for $S_8$ across the three summary statistics, while $w$ is comparatively less well calibrated.
The TARP results also show relatively small deviations across all three summary statistics, indicating that the joint posteriors are reasonably well calibrated.

\section{Scale cuts}
\label{sect:appx_cut}
In weak-lensing analyses, scale cuts are commonly used to mitigate the impact of insufficiently modelled physical processes and potential systematic contaminants, such as source clustering, baryonic feedback, and additive biases arising from PSF errors \citep{Gatti2024_sbi}.
As shown in Fig.~\ref{fig04:data_vectors_subset}, the impact of source clustering on our summary statistics is particularly pronounced on small scales.
We therefore investigate the effect of excluding angular scales below $6.88\,\mathrm{arcmin}$ in the third-order moments.
For the wavelet statistics, we instead exclude the wavelet coefficient map at the smallest nominal angular scale of $3.44'$.
We then repeat the cosmological inference using the remaining scales and compare the results with those obtained from the full scale range.

\begin{figure}[ht!]
\centering
\includegraphics[width=85mm]{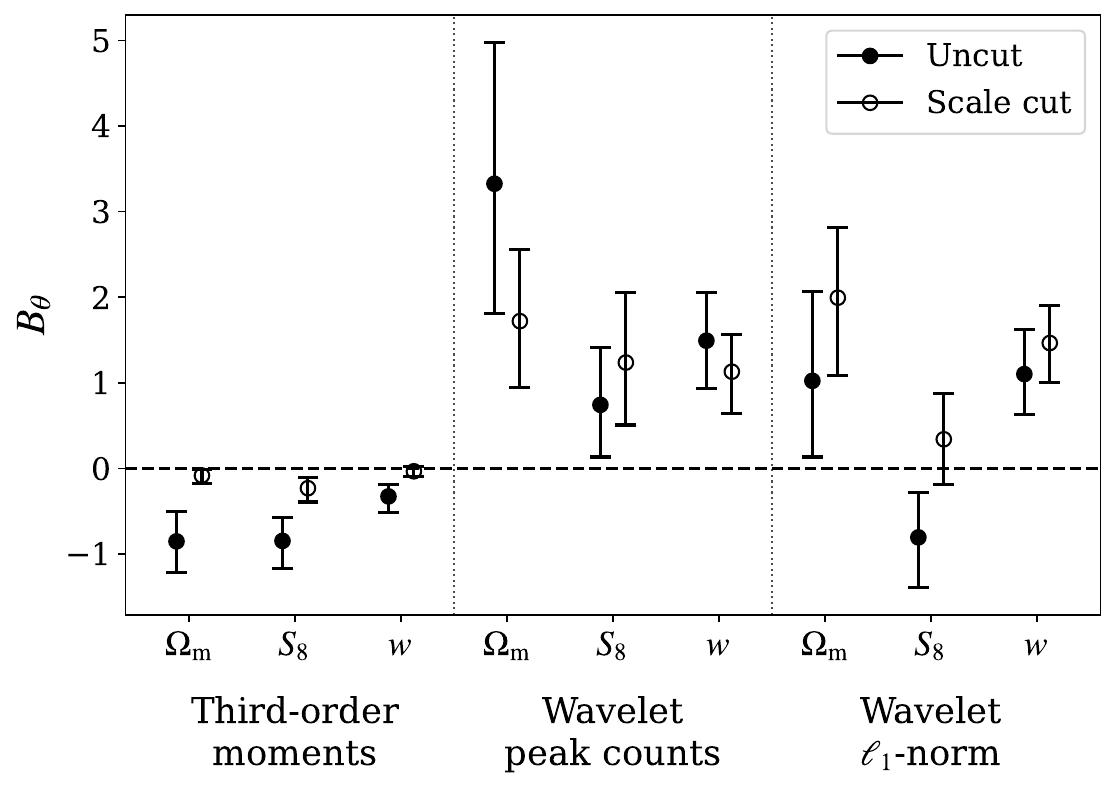}
    \caption{Source clustering bias for $\Omega_{\mathrm{m}}$, $S_8$, and $w$ across the three summary statistics. Circles and squares indicate patch-level and split-level results, respectively.}
\label{figb01:sc_bias_cut}
\end{figure}

\begin{figure}[ht!]
\centering
\includegraphics[width=85mm]{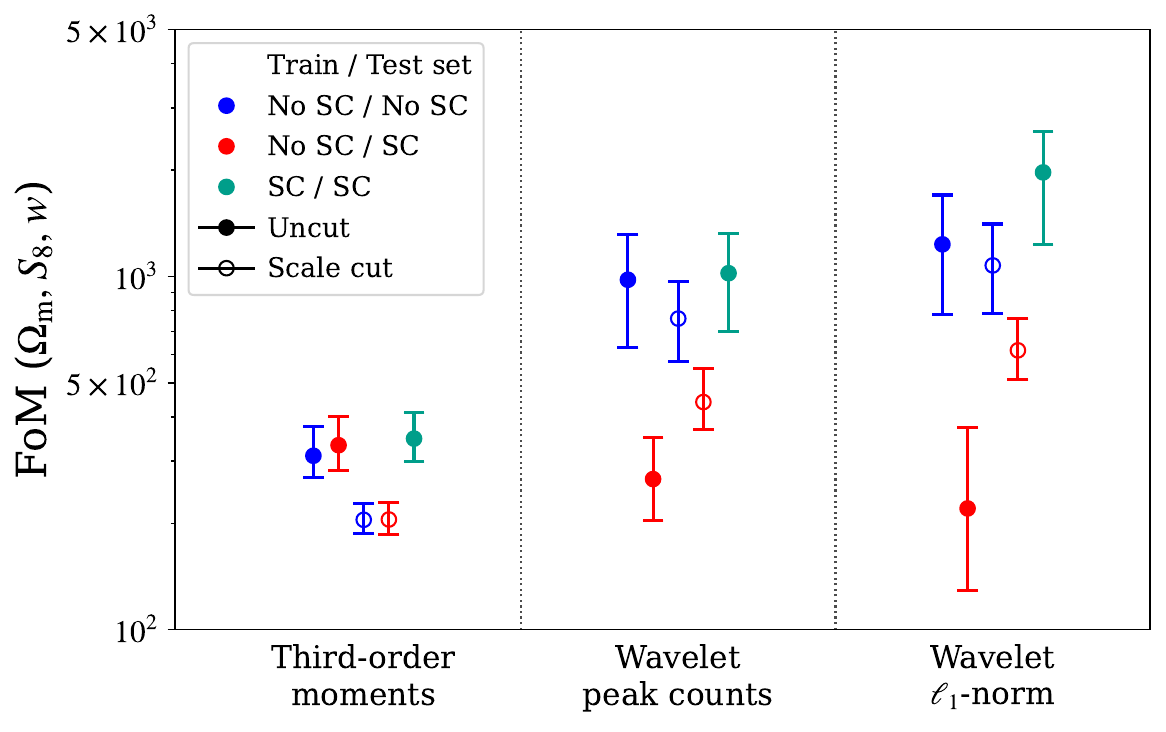}
    \caption{Figure of merit for the three summary statistics, considering the $\Omega_\mathrm{m}$, $S_8$, and $w$ parameter combination. Blue, red, and teal correspond to No SC / No SC, No SC / SC, and SC / SC, respectively, where the first and second entries refer to the training and test sets.}
\label{figb02:fom_sc_cut}
\end{figure}

We focus on patch-level results here, as the split-level results primarily reflect the robustness to the choice of training and test sets.
Fig.~\ref{figb01:sc_bias_cut} shows the source clustering bias after applying scale cuts.
The impact of scale cuts varies across summary statistics and cosmological parameters.
For the third-order moments, scale cuts substantially reduce the bias, bringing it close to zero.
In contrast, the bias slightly increases for $S_8$ with wavelet peak counts and for $\Omega_{\mathrm{m}}$ with the $\ell_1$-norm.
This increase can be understood in terms of the competing effects of scale cuts.
While excluding small scales reduces the impact of mismodeling, the source-clustering signal persists on larger scales in the wavelet maps.
As shown in Fig.~\ref{figb02:fom_sc_cut}, scale cuts mitigate the FoM degradation caused by mismodeling, resulting in smaller posterior uncertainties and, consequently, a smaller denominator in the bias estimator of Eq.~(\ref{eq14:bias_estimator}).
This leads to a larger estimated bias for the two cases mentioned above.

%%%%%%%%%%%%%%%%%%%%%%%%%%%%%%%%%%%%%%%%%%%%%%%%%%%%%%%%%%%%%%

\FloatBarrier %\usepackage{placeins}
\clearpage

\end{appendix}
\end{document}